\documentclass[prl,twocolumn,amsmath,amssymb,superscriptaddress]{revtex4}

\usepackage{graphicx}

\newcommand{\la}{\label}
\newcommand{\be}{\begin{equation}}
\newcommand{\ee}{\end{equation}}
\newcommand{\bea}{\begin{eqnarray}}
\newcommand{\eea}{\end{eqnarray}}
\newcommand{\p}{\partial}
\newcommand{\ba}{\begin{align}}
\newcommand{\ea}{\end{align}}
\newcommand{\1}{\frac{1}{2}}

\begin{document}
\title{Quantum Hydrodynamics, Quantum Benjamin-Ono Equation, and Calogero Model}
\author{Alexander G. Abanov}
\affiliation{Department of Physics and Astronomy,
Stony Brook University,  Stony Brook, NY 11794}
\author{Paul B. Wiegmann}
\affiliation{James Franck Institute
of the University of Chicago, 
5640 S.Ellis Avenue, Chicago, IL 60637,
\altaffiliation[Also at ]{Landau Institute of Theoretical Physics.}}

\date{\today}

\begin{abstract}
Collective field theory for Calogero model represents particles with fractional statistics in terms of hydrodynamic modes -- density and velocity fields. We show that the quantum hydrodynamics of this model  can be written as a single evolution equation on a real holomorphic Bose field -- quantum integrable  Benjamin-Ono equation. It renders tools of integrable systems to studies of nonlinear dynamics of 1D quantum liquids.
\end{abstract}
\maketitle


\paragraph*{1. Introduction.}

Calogero-Sutherland model
occupies a special place in 1D quantum
physics. A singular  interaction (an inverse square potential)
represents a fractional statistics of particles  
(for a review see 
e.g., \cite{1999-Polychronakos}).
The eigenfunctions of the model are neither symmetric nor antisymmetric. They
have a form $(\Delta(x))^\lambda J(x)$, where  $\Delta$ is a 
Vandermonde determinant
and $J$  are   symmetric polynomials (Jack polynomials).
Although the problem does not possess a conformal invariance, it is 
intrinsically related to a  boundary conformal field theory (CFT) 
\cite{Awata} with a central charge
$c=1-24\alpha_{0}^{2}$, where
$2\alpha_{0}=\sqrt{\lambda}-1/\sqrt{\lambda}$.
Here we focus on a rational degeneration of the model -- the
Calogero model (CM)
\be
 \label{1}
	{\cal H} = \frac{1}{2}\sum_{j=1}^{N}p_j^{2}
	+\1\sum_{j,k=1; j\neq k}^{N}
	\frac{\lambda(\lambda-1)}{(x_{j}-x_{k})^{2}}.
\ee
One might also add a harmonic potential $\1\omega^2
\sum_i x_i^2$ to prevent particles from running to infinity. Most of
the formulas below are simple to generalize to  the chiral sector of 
the trigonometric
version of the model.

A productive approach to Calogero model in the limit
of an infinite number of particles is a collective field theory
developed by Sakita and Jevicki in Refs. 
\cite{JevickiSakita, Jevicki-1992} and extended to CM
in Refs. \cite{Awata, AJL-1983}. In this approach an entire quantum
theory  with all the richness of an operator content,  is represented as
a quantum dissipationless hydrodynamics
\cite{Landau-1941}, i.e., solely in terms of  the density
$\rho(x)=\sum_i\delta(x-x_i)$ and the velocity operators
$$
\left[\rho(x),v(y)\right] = -i\delta'(x-y).
$$
Hydrodynamic approach is
especially useful if one is interested  in  the evolution of  smooth wave
packets,  larger than an inter-particle distance. Original particles 
apear in this
approach  as solitons of nonlinear fields.

In this letter we show that  the quantum  hydrodynamics (aka  bosonization,
or  collective field theory) of Calogero model is equivalent to a
\textit{quantum}
version of an integrable {\it Benjamin-Ono equation on the double} (QBO). 
The latter is a
minor  generalization of the conventional {\it
Benjamin-Ono}
equation (BO) arising in the hydrodynamics of stratified fluids
\cite{AblowitzClarkson-book}.
A connection of the traditional BO to a complexified version of a classical 
Calogero model is  known \cite{CLP-1979}.
A pair of  coupled classical BO-type equations has been obtained  by  A.
Jevicki  for a  long-wave description of free fermions ($\lambda=1$)
\cite{Jevicki-1992}.


\paragraph{2. Quantum hydrodynamics and non-linear bosonization.}
Collective  or hydrodynamic approach is a nonlinear version of the 
bosonization procedure \cite{Stone-book} - a popular method in low
dimensional physics. In this approach, a fermionic spectrum is not
linearized at the Fermi level. Therefore, an asymmetry
of the particle-hole spectrum  - a dispersion of hydrodynamics modes - is
not neglected. The  price  is a nonlinear and dispersive
character of  the hydrodynamics.  The gain is a realm of nonlinear
phenomena missed in a linear approximation.
Before proceeding to the Quantum BO equation on the double (QBO), we 
start with the quantum hydrodynamics of
free fermions.

Quantum hydrodynamics  of free fermions with
quadratic dispersion
($\lambda=1$) can be  obtained using conventional 
``bosonization''.
Writing a fermion
$\psi(x) =  e^{i\varphi_{R}(x)}+e^{i\varphi_{L}(x)}$ as a 
superposition of left and right
chiral fields $[\partial_x \varphi_{L,R}(x),\,\varphi_{L,R}(0)]=\mp
i\delta(x)$ one obtains
\be
   \la{3}
	{2\pi (n+1)}\psi^\dag(-i\partial_{x})^n \psi
   	\sim :{(\p_x\varphi_R)^{n+1}-(\p_x\varphi_L)^{n+1}}:
\ee
modulo full derivatives. \textit{Here and below we assume normal ordering}.
The first three moments are density
$\rho=\frac{1}{2\pi}\partial_{x}(\varphi_{R}-\varphi_{L})$,
current
$j=\rho
v=\frac{1}{4\pi}((\partial_{x}\varphi_R)^2-(\partial_{x}\varphi_L)^2)$,
and Hamiltonian
density
\be
   \label{HFFb}
	H =  \frac{1}{12\pi}
((\partial_{x}\varphi_{R})^{3}-(\partial_{x}\varphi_{L})^{3})
=\frac{1}{2} \rho v^{2}+\frac{\pi^{2}}{6}\rho^{3}.
\ee
The operator
$v =\1 \partial_{x}(\varphi_{R}+\varphi_{L})$ is a velocity operator.
The meaning of the terms in the energy (\ref{HFFb}) is clear. The
``kinetic energy''
reflects  the Galilean invariance of (\ref{1}), the ``potential
energy'' reflects the fermionic
statistics: $(\pi^2/6)\rho^3$ is an energy density of a Fermi gas.
This Hamiltonian takes  into account all long-wave
    correlation properties of fermions, missing the physics at $2k_F$
 (e.g., Friedel
oscillations),  where the chiral sectors interact. Linearization at
the equilibrium
density
$\rho_0=k_F/\pi$  leads to  a familiar linear
bosonization 
$H\approx
\frac{\rho_0}{4}((\p_{x}\varphi_R)^2+(\p_{x}\varphi_L)^2)=\frac{\rho_0}{2}
(v^2+\pi^2(\rho-\rho_0)^2).$
This  approximation misses
an interaction and dispersion of hydrodynamics modes, essential in many
applications.

The  Hamiltonian (\ref{HFFb}) generates operator equations of quantum
hydrodynamics \cite{Landau-1941}
\bea
   \la{5}
	{\rm continuity\,equation}:&\qquad\dot \rho+\p_x(\rho v)
   	&= 0,\\
   \la{6}
	{\rm Euler\, equation}:&\qquad\dot v+\p_x (\1 v^2+w)
	&= 0,
\eea
where $\quad w=\delta(\rho\epsilon)/\delta\rho$ is the quantum enthalpy
and $\epsilon=(\pi^2/6)\rho^2$ is the energy of a filled Fermi sea 
per particle.

\paragraph{3. Holomorphic Bose field.}

Both continuity equation (\ref{5}) and Euler equation (\ref{6}) for free
fermions can be compactly written in terms of a single holomorphic
field $u(z,t)$. It is a real field obeying the Schwarz reflection symmetry
$u(z)=\overline{u(\bar z)}$. This field glues together  left and
right components of the Bose field.
\begin{figure}[h111]
	\begin{center}
     \includegraphics[width=1.5in]{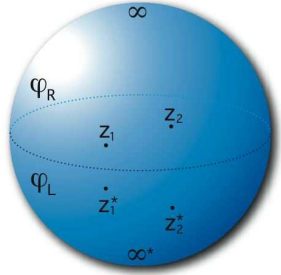}
     \caption{\label{fig:double} The double. Left  and
right  sectors are glued together along the branching cut (equator) 
-- a segment of the real axis supporting a particle density. Excitations are 
symmetrically located in a complex plane.
}
	\end{center}
\end{figure}
Let us pass to an imaginary time $t\to -it$, $v\to iv$, and treat 1D
coordinate $x$  as a real axis of a complex plane
$z$.  Then right and left currents $\pm \p_{x}\varphi_{R,L}=\pi \rho 
\pm iv$ are complex
conjugates and may be treated as boundary values of an
analytic field  in a domain adjacent to the real axis
$u(x\pm i\,0,t)=v\pm i\pi \rho=-i\p_{x}\varphi_{R,L}$.
Then the hydrodynamic equations  (\ref{5},\ref{6})
are seen as boundary values ($z=x\pm i\,0$) of the quantum
Hopf-Burgers equation
\be\la{7}
	\p_t u+u\p_zu=0.
\ee
This is a nonlinear bosonization of free fermions.

The analytical structure allowing to treat the right 
current as an analytical
continuation of the left one across the boundary (the real axis $x$)
extends to the Calogero model. There, as we argue below,
the boundary values of  the field $u$ are
\be
   \la{8}
	u(x\pm i\,0)=(1/\sqrt{\lambda}) (v \pm i\lambda\pi\rho)
 +\alpha_{0} \p_x\log \rho,
\ee
The field $u$ consists of  positive and negative parts
$$u=u^{-}(z)+u^{+}(z).$$
The negative part  is a Cauchy transform of the density
$$
	u^{-}(z)=\sqrt{\lambda} \int dx\,\frac{\rho(x)}{x-z}.
$$
It is an analytic function in a  plane cut by a segment of a real axis
supporting the density. Its
   boundary values $u^{-}(x\pm i\,0)=\pi\sqrt{\lambda}(
\rho^{H}\pm i\,\rho)$, where
$\rho^{H}=(1/\pi) P\int dx'\, \rho(x')/(x'-x)$
is a Hilbert transform of the
density.
A positive part -- $u^{+}(z)$ is an analytic continuation of the real function
$u^{+}(x)=(1/\sqrt{\lambda}(v-\lambda\pi\rho^H)+\alpha_{0}\p_x\log\rho$
from the real axis. It
is analytic only in the vicinity of the real axis, having singularities
elsewhere.

Let us mark two points, one inside, another outside of the domain, 
where both $u^{+}$ and $u^{-}$ are analytic,
and expand the fields around these points. For simplicity let us
talk about the rational case. The natural points are
$z=0$ and $z=\infty$. We have
\be
 \la{100}
	u^{-}(z) = -\sum_{n\ge 0} a_{n}z^{-n-1},\;\;
   	u^{+}(z) =-\sum_{n\ge 1} a_{-n}z^{n-1}.
\ee
The modes $a_{n}$ are  real and Hermitian. Positive modes are moments
of density (collective variables)
$a_n=\sqrt{\lambda}\int x^n\rho(x) dx$ and negative modes are
canonically conjugated to positive ones
$[a_{n},\,a_{-m}]=-n\delta_{n,m}$.   The zero-mode is the number of
particles $a_0=\sqrt{\lambda} N$. It does
not change in time. The field $u(z)$ is a current of a real canonic Bose field
$\varphi=\varphi^{+}+\varphi^{-}$, where
$\p_{z}\varphi^{\pm}(z)=u^{\pm}(z)$.
%
\indent Let us notice a difference  between the Bose field (\ref{100}) and a Bose
field  used in conventional bosonization of free fermions with a linearized
spectrum. In the former, the time $t$ is not a part of a $z$-plane, but describes an  evolution of a \textit{real} holomorphic field $u(z,t)$. In the latter, time and space form a complex coordinate $x+it$ of the
radial quantization. Complex $z$-plane has an analytical structure 
similar to ones used in boundary CFT and is sometime referred to 
as {\it (Schottky) double}.

\paragraph{4. Quantum Benjamin-Ono equation} 
is defined as
\be
	\p_t u+\partial_z\left(\1 u^{2} 
+\alpha_{0}\partial_{z}(u^{+}-u^{-})\right)=0,
   \label{QBO}
\ee
where we rescaled time  $t\to t/\sqrt{\lambda}$. As the
Hopf-Burgers equation (\ref{7}), it is a quantum  equation acting in
the Fock space spanned by modes $a_n$. QBO is integrable. It has infinite number of conserved integrals.
The first four are $u$ itself, momentum density  
$T(z)=u^{2}/2+i\alpha_0\p(u^{+}-u^{-})$, the Hamiltonian
\be
	H=
   	\oint \frac{dz}{4\pi i}\,
	\left[\frac{1}{3}u^{3}
	+\alpha_{0} u\partial_{z}(u^{+}-u^{-}) \right],
   \label{12}
\ee
and the fourth integral
\be
	I_{4}=
   	\oint \frac{dz}{2\pi i}\,
	\left[\frac{u^{4}}{4} -\frac{3\alpha_{0}}{2} u^{2}\partial_{z}(u^{+}-u^{-})
	+2\alpha_{0}^{2} (\partial_{z}u)^{2} \right],
   \label{4-int}
\ee
Here the integration goes around the real axis.
Below we show that (\ref{QBO},\ref{12}) are equivalent to the
CM (\ref{1}).

The classical BO
equation originally appeared in hydrodynamics of stratified fluids. There the real
harmonic function $u(x,y)$ on  a half plane evolves following a
time-dependent boundary
condition on the real axis \cite{AblowitzClarkson-book}
\be
  \la{OBO}
  	\p_t u+u\partial_xu 
+\partial_{x}^{2} u^{H}=0.
\ee
The Hilbert transform  $u^{H}
=(1/\pi) P\int dx'\, u(x')/(x'-x)$ can be seen as $u^{H}=i(u^{+}-u^{-})$
with respect to the one marked point at infinity.
Contrary,  our  BO-equation  has two the marked points $z=0$ and $z=\infty$.
The term $(u^{+}-u^{-})=-iu^{H}_{C}$ in (\ref{QBO}) can be seen as a
Hilbert transform with respect to the contour $C$ embracing a
support of the density (real axis)\linebreak
$u^{H}_{C}(z\in C)=\frac{1}{\pi}P\oint_{C}\frac{u(z')}{z'-z}\,dz'.$
This difference is not an
obstacle to an almost automatic extension of known solutions
of the traditional BO equation. Remarkably, this
extension  carries over to the quantum case as well.

\paragraph{5. Quantum Hirota equation.}

Typically, integrable equations can be written in the bilinear (Hirota's)
form (see for a review \cite{Matsuno-book, Zabrodin-1997}).  Apparently, 
bilinear forms exist also for quantum integrable
equations.
Let us introduce a vertex operator
$\Psi_{\sigma}= \exp(-\sigma\varphi)$,
where $\sigma=\sqrt{\lambda}$ or $\sigma=-1/\sqrt{\lambda}$, and consider its
positive and negative parts
\be
   \la{vertex}
   	\psi^{-} = e^{\sigma\varphi_{-}},
	\quad
   	\psi^{+} = e^{-\sigma\varphi_{+}}.
\ee
The bilinear quantum Hirota equation
\be
   \la{14}
	D_t\,\psi^-\cdot\psi^+ =-\frac{1}{2\sigma} D_z^2\,\psi^-\cdot\psi^+,
\ee
is an equivalent form of QBO (\ref{QBO}).
Hirota's derivatives are defined as
$D_x^na\cdot b=(\p_x-\p_{x'})^na(x)b(x')|_{x=x'}$.  
The proof is straightforward.

The  Hirota equation (\ref{14}) is a compact form of 
OPE (\ref{3}), extended for Calogero
model. The resemblance to the
Schr\"odinger equation further clarifies the  origin of QBO.
It is especially transparent for   free
fermions $\lambda=1$. There, the vertex operator $\Psi_{\sigma}$
is a bosonized form of a
fermion creation ($\sigma=1$) and anihilation ($\sigma=-1$) 
operators. 
The Schr\"odinger equation
for free fermions $\p_t\Psi=\1\p^2_z\Psi$ transforms to Hirota equation
(\ref{14}) by an elementary algebra.
A known duality between particles and holes $a_n\to a_{-n}$ under a
transformation
$\sqrt{\lambda} \to -1/\sqrt{\lambda}$ is  especially transparent in the bilinear form.

\paragraph{6. Quantum hydrodynamics and QBO-equation.}

Relation to hydrodynamics is obtained by specializing   BO-equation on the real
axis. Computing the jump (discontinuity) of the l.h.s.~of (\ref{QBO})
across the real axis we obtain the continuity equation (\ref{5}) with
the current
$j=\rho v=(T(x+i\,0)-T(x-i\,0))/(2\pi i\sqrt{\lambda})$. It gives the announced
connection (\ref{8}) between the field
$u$ and velocity. The sum of BO equations taken on two sides of the
real axis gives the Euler equation (\ref{6}) with the energy density
\be
   \la{enden}
   	\epsilon= \frac{1}{6}(\pi \lambda\rho)^{2}
	+\frac{\alpha_0^{2}}{2}(\partial_{x}\ln\rho)^{2}
	+\pi\alpha_0\lambda \partial_{x}\rho_{H}.
\ee
Eqs. (\ref{5},\ref{6},\ref{enden})  have been derived in Ref.
\cite{AJL-1983} within the collective field
theory approach.

\paragraph{7. Pole dynamics of quantum Benjamin-Ono equation and
Calogero  model.} Let us now show that QBO is equivalent to
the Calogero model.

An important class of solutions of classical integrable
nonlinear equations is given by a moving  pole ansatz. These  solutions
have  a form  $u=\sum_{i=1}^{N} f(z-x_i(t))$, where $f(x)$ is an elliptic
    function, or its trigonometric or rational degeneration
(see, e.g., \cite{AblowitzClarkson-book}). Then
nonlinear equation is reduced to a  certain, also integrable, many body
problem of $N$ particles with coordinates $x_i$. In most cases it 
is the Calogero model and its decendents. 
In Ref. \cite{CLP-1979}  similar fact has been noticed for the classical BO-
equation (\ref{OBO}). The complex poles $x_{i}(t)$ of a  pole ansatz
$u(x,t)= -\sum_i(\frac{i\lambda}{x-x_i}-\frac{i\lambda}{x-\bar x_i})$
move according to a  complexified classical Calogero equation
\be
    \la{177}
	\ddot x_i=2\sum_{j,j\neq i} \frac{\lambda^2}{(x_i-x_j)^{3}}
\ee
under the  condition that ${\rm Im}\, x_{i}>0$ at an initial time.

Our version (\ref{CBO}) of the classical BO equation  
produces a genuine Calogero model where
coordinates of particles (poles) are real. 
Remarkably, the  pole ansatz carries over to the quantum case. Pole ansatz  is
perhaps a
simplest way to  see the equivalence between CM and QBO.

The algebra is straightforward and can be carried out for both QBO 
and quantum Hirota equation. We briefly  describe the main
steps starting from the QBO (\ref{QBO}). 

We look for a solution  in the form
\be
   \la{17}
	u^{-}(z,t)=-\sqrt{\lambda}\sum_{i=1}^N\frac{1}{z-x_i(t)},\quad
x_i=\rm{Real}
\ee
and show that the dynamics of  the  poles  $x_i$ obey quantum
Calogero model with $N$ particles.  The residue in (\ref{17})  is
controlled by the asymptote at infinity $u^{-}(z)\sim (-\sqrt{\lambda}N)/z$. 
At the pole ansatz the vertex operator (\ref{vertex}) at $\sigma=-1/\sqrt{\lambda}$   has simple zeros at  positions 
of particles  $\psi^-=\prod_j(z-x_j)$.

We plug (\ref{17}) into the negative part of (\ref{QBO}) and assume  $\left[u^{+}(x_{i}), x_{j}\right]=-\delta_{ij}/\sqrt{\lambda}$, which yield canonical commutation relations for $u^{\pm}$ as  $N\to\infty$.
Using formulas
$(u^{-}u^{+})^{+}=-\sum_j\frac{\sqrt{\lambda}}{z-x_j}(u^{+}(z)-u^{+}(x_j))$ and
$(u^{-}u^{+})^{-}=-\sum_j\frac{\sqrt{\lambda}}{z-x_j}u^{+}(x_j)$, we 
obtain that a
rational function
\bea
	&&
	\frac{1}{z-x_j} \dot{x}_j\frac{1}{z-x_j}
	+ \sqrt{\lambda}\sum_{k,k\neq j}\frac{1}{(x_j-x_k)(z-x_j)^2} \nonumber
 \\
	&& 
+\frac{1}{\sqrt{\lambda}}\frac{1}{(z-x_j)^{3}}-\frac{1}{(z-x_j)^2}u^+(x_j)
 	=0\la{188}
\eea
vanishes. Both second and third order poles vanish if $\dot{x}_{j} 
=p_{j}$ with $\left[p_{i}, x_{j}\right]=-\delta_{ij}/\sqrt{\lambda}$ 
and
\be
 \la{18}
 	p_{j} =u^{+}(x_{j}) 
-\sqrt{\lambda}\sum_{k,k\neq j}\frac{1}{x_{j}-x_{k}}.
\ee
Eq. (\ref{18}) is a finite $N$ version of the equation (\ref{8}). 
Differentiating (\ref{18}) over time and using the positive part of 
(\ref{QBO}), we obtain after some work Eq.(\ref{177}) with 
$\lambda^{2}\to \lambda(\lambda-1)$ (where time is scaled back to the 
original real time).

Few comments are in order: (i) The pole ansatz (\ref{17})
represents   positive  modes
$a_n=\sqrt{\lambda}\sum_jx_j^n$. Negative
modes are represented in terms of momenta and coordinates of
particles  through (\ref{18}). Canonical commutation relations
between positive and negative modes appear only in the field theory
limit. For a finite number of particles they are replaced by
the conditions  $[u^{+}(x_i),\, x_j]=-\delta_{ij}/\sqrt{\lambda}$
used in the cancelation of poles; (ii)
Similar  manipulations in the classical case produce
Calogero model with interaction $\lambda^2(x_i-x_j)^{-2}$. A  quantum
shift $\lambda^{2}\to \lambda(\lambda-1)$ comes from pulling momentum 
operator $p_j$ to
the right in the first term of (\ref{188});
(iii) A trigonometric  ansatz $u^{-}(z,t)=-\sqrt{\lambda}k\sum_i \cot(k(z-x_i(t))$ leads
to the Calogero-Sutherland (trigonometric)  model.

\paragraph{8. Solitons and integrability}
Let us now consider the large $N$ limit and assume that there is a
non-zero background density of particles. Then,  alternatively, one looks for
the pole ansatz  for  $u^{+}$ with a finite
number of poles symmetrically located with
respect to the real axis (see Fig.\ref{fig:double})
\be
 \la{21}
	u^{+}(z,t)=\sqrt{\lambda}\sum_{i=1}^n
	\left(\frac{1}{z-z_i}+\frac{1}{z-z^*_i}\right).
\ee
In this case the vertex operator (\ref{vertex}) at 
$\sigma=-1/\sqrt{\lambda}$ is $\psi^+(z)=\prod_i(z-z_i)$ and has zeros at
positions of poles. \cite{Awata} These poles  also move according to quantum
Calogero model. They  correspond to excitations
of a system with  a continuous density.
Accordingly $u^{-}(z)$ has a branch cut on the real axis. Solutions
(states) with well-separated poles $z_{i}$ of $u_+(z,t)$ are quantum analogs of
solitons known in classical nonlinear equations. A semiclassical one-soliton solution  of the Euler equation  (\ref{6}) studied in Ref.
\cite{Jevicki-1992,Polychronakos-1995,AndricBardekJonke-1995}, corresponds to a single pole in (\ref{21}) and can be found from (\ref{8}). 

Existence of solitons signals that the QBO is integrable.
Indeed, the known Lax representation for the classical BO equation
(\ref{OBO})   extends to the
quantum BO equation on the double (\ref{QBO}) with minor changes. In
particular, the pole ansatz converts Lax operators of the quantum BO 
equation to known Lax
operators of quantum  Calogero model. The same  is true for the conserved
integrals, B\"aklund transformation, etc.  In a standard way these
tools provide multi-soliton, periodic, and shock-wave solutions for
quantum hydrodynamics.

\paragraph{10. Semiclassical limit of BO equation and nonlinear transport
in quantum systems.} 

Classical BO equation may be useful in
studying nonlinear effects in  transport in low-dimensional quantum
systems. It describes a semiclassical  evolution of long wave packets of
density. 
The semiclassical   limit is subtle and requires certain care. In nonlinear quantum 
equation (\ref{QBO}) a simple
replacement of a quantum field
$u$, by classical fields is incorrect. The correct result
is obtained by shifting  $\alpha_{0}\to \sqrt\lambda/2$ in (\ref{QBO})\linebreak
\be
	\p_t u+\partial_z\left(\1 u^{2}
+(\sqrt{\lambda}/2)\partial_{z}(u^{+} -u^{-})\right)=0.
   \label{CBO}
\ee
In particular, the dispersive (last) term in  (\ref{CBO}) does not vanish for free fermions at
 $\lambda=1$, while it vanishes  in the  quantum version (\ref{7}).
The equation (\ref{CBO}) (or rather a corresponding classical action)   must be
understood as a leading term of the gradient expansion of the effective
action, with the first order correction due to zero fluctuations. It can be
 summurized as
$ \langle u^2\rangle-\langle u\rangle^2
=\sqrt{\lambda}\,\p_z\langle u^+-u^-\rangle$, where we average over the 
coherent state with a given density. The correction ammounts to the shift
 $\lambda(\lambda-1)\to \lambda^2$. One can also obtain this
quantum correction by comparing the classical
(\ref{177}) and quantum (\ref{1}) CM obtained as pole ansatzes of classical 
and quantum  BO equations. The classical pole asatz (\ref{17}) leads
to the classical Calogero equation (\ref{177}) only when it is plugged into
(\ref{CBO}) rather than into
a naive classical version of (\ref{QBO}).
The classical dynamics
of long wave packets depends on $\lambda$ only through scales. It is 
described by the classical BO on the double, and is the same 
for free fermions and ``anyons''.

The appearance of the dispersion term (the last term in (\ref{CBO})) for free
fermions has been pointed out  by A. Jevicki \cite{Jevicki-1992}. It has far
reaching consequences. As it is well known, a dispersionless limit of
non-linear equations is singular \cite{book}. Almost any smooth initial configuration
$u(x,t)$ evolving according to a
classical Hopf-Burgers equation (\ref{7}) (a  dispersionless limit of
(\ref{CBO})) develops an unphysical shock wave in finite time. The
dispersion being initially small becomes a crucial factor  and can not be neglected. A discussion of the role of
dispersion in electronic transport, and a number of other interesting
topics, like quantum integrability, relation to perturbed conformal field
theory etc.,  is outside of the scope of this letter. We plan to 
address some of these
issues  elsewhere.

We have  benefited from discussions with  O. Agam, E. Bettelheim, A. Mirlin, L. Takhtajan, A. Tsvelik, and
A. Zabrodin. Our special thanks to I. Gruzberg, D. Gutman and R.
Teodorescu  for help, and
contributions. P.W. was supported by the NSF
MRSEC Program under DMR-0213745 and NSF DMR-0220198. P.W. acknowledges
support  by the Humboldt foundation.
The work of AGA was supported by the NSF under the grant DMR-0348358 and by the Theory Institute  at BNL.



\end{document}